\documentclass[a4paper, 12pt]{article}
\usepackage[utf8]{inputenc}
\usepackage{color}
\usepackage{hyperref} 
 \usepackage{latexsym}
  \usepackage{amssymb}
  \usepackage{graphicx}
   \DeclareGraphicsExtensions{.jpg .png .gif .pdf .pdf}
\usepackage[T1]{fontenc}
 \usepackage[english]{babel}
  \usepackage{}
  \usepackage{amstext}
  \usepackage{amsmath}
  \usepackage{times} 
  \usepackage{amsfonts}
 \date{}

\title{Melnikov chaos in a modified Rayleigh-Duffing oscillator  with $ \phi^6$ potential}

\author{C. H. Miwadinou\footnote{clement.miwadinou@imsp-uac.org, hodevewan@yahoo.fr}, A. V. Monwanou\footnote{Author to whom correspondence 
should be addressed: movins2008@yahoo.fr, vincent.monwanou@imsp-uac.org},  L. A. Hinvi
\footnote{laurent.hinvi@imsp-uac.org},  A. A. Koukpemedji\footnote{kaudranus2000@gmail.com}, \\ C.
 Ainamon\footnote{ainamoncyrille@yahoo.fr} and J. B. Chabi Orou\footnote{ jchabi@yahoo.fr}} 

\begin{document}

\maketitle Institut de Math\'ematiques et de Sciences Physiques, Universit\'e d'Abomey-Calavi, BP: 613 Porto Novo, B\'enin
\begin{abstract}
The chaotic behavior of the modified Rayleigh-Duffing oscillator with $ \phi^6$ potential and external excitation which 
modeles ship rolling motions  are investigated both analytically
and numerically. Melnikov method is applied and the conditions for the existence of homoclinic and heteroclinic chaos are obtained.
The effects of nonlinear damping on roll motion of ships are analyzed in detail. As it is known, nonlinear roll damping is a very important parameter
in estimating ship reponses. The predictions are tested  numerical simulations based on the basin of attraction. We conclude that certains quadratic
 damping effects are contrary to cubic damping effect.
\end{abstract}

{\bf keywords} Modified Rayleigh-Duffing oscillator, Melnikov criterion, nonlinear damping effect, 
ship rolling, basin of attraction

\section{Introduction}
 The analysis of nonlinear dynamic systems is now a major theme in both  academic and industrial perspective and touches many areas, such as 
hydrodynamics, aerospace, civil engineering, transport, musical acoustics, nuclear engineering and others
 [Enjieu Kadji and Nana Nbendjo, 2012; Hayashi, 1964; Nana Nbendjo et al., 2007; Nayfeh and Mook, 1979; Tchoukuegno, 2002; Tchoukuegno, 2003; 
Yamapi, 2003; Yamapi, 2007]. The first works date from the nineteenth century
 including Poincar\'e, but currently it is experiencing a resurgence of interest due to the need to optimize, streamline structures commonly used and
 subjected to significant levels of excitement, or control the instabilities oscillations. Modeling and study of the behavior of a physical system is a
 major problem even complicated in case the present system nonlinearities. Much of the discussion in the physics and engineering literature concerning 
damped oscillations, focuses on systems subject to viscous damping, that is, damping proportional to the velocity, even though viscous damping occurs 
rarely in physical systems. Phenomenological models describing some type of nonlinear dissipation have been used in some applied sciences such as ship
 dynamics [Bikdash et al., 1994; Falzarano et al., 1992], where a particular interest has deserved the role played by different damping mechanisms in the
formulation of ship stability criteria, and vibration engineering  [Ravindra and Mallik, 1994a; Ravindra and Mallik, 1994b]. Damping in certain
 applied systems plays an important role, since it may be used to suppress large amplitude oscillations or various instabilities, and it can be also used 

as a control mechanism [Litak et al., 2009; Miwadinou et al., 2015; Rand et al., 2000; Sanju\'an, 1999; Soliman and Thompson, 1992,
Taylan, 2000].

This paper is concerned with the appearance of homoclinic and heteroclinic instabilities and chaos in a triple-well modified Rayleigh-Duffing
oscillator. Many problems in physics, chemistry, biology, Engineering,  etc., are related Rayleigh-Duffing oscillator.
[Miwadinou et al., 2014; Miwadinou et al., 2015; Siewe Siewe et al., 2009]. 
The dynamics of Rayleigh-Duffing oscillator has been investigated widely in these years
[Miwadinou,]. For example, in their work, Siewe Siewe et al. [2006]  studied the nonlinear response and suppression of chaos
 by weak harmonic  perturbation inside a triple well $\phi^6-$Rayleigh oscillator combined to parametric excitations. In [Siewe Siewe et al., 2009], 
the authors focussed their analysis on the occurrence of chaos in a parametrically driven extended Rayleigh oscillator with three-well potential 
modeled by:
\begin{eqnarray}
 \ddot{x} -\mu (1-\dot{x}^2)\dot{x}+(1-\gamma\cos\omega t)(x+\alpha x^3+\beta x^5)=F\cos\omega t. \label{eq.001}
\end{eqnarray}

 On the other hand, the Rayleigh-Duffing oscillator is used to model the roll motion of ships. 
 The nonlinear ship rolling response  generally is given by [Francescutto and  Contento, 1999; Holappa et al., 1999; Scolan, 1999; Taylan, 2000] 
\begin{eqnarray}
 &&\ddot{x}+2\mu \dot{x}+\beta_1|x|\dot{x}+\beta_2\dot{x}|\dot{x}|+\delta_1x^2\dot{x}+\delta_2\dot{x}^3+
+...+\omega_0^2x+a_3x^3+a_5x^5+\cr
&&...=F\cos\omega t, \label{eq.01000}
\end{eqnarray}
where $\omega_0$ and $\omega$  are internal and external frequencies respectively; $\mu$, $\beta_i$ and $\delta_i$ are linear, quadratic nonlinear
and cubic nonlinear damping coefficients respectively and $a_i$ are restoring coefficients. $F$ is the external excitation amplitude.
In these cases,  Francescutto and  Contento [1999]  used experimental results and parameter identification technique to study bifurcations in ship
rolling modeled as :
\begin{eqnarray}
 \ddot{x}+2\mu \dot{x}+\delta_2\dot{x}^3+\omega_0^2x+\sum_{j=1}^{11} a_jx^j=F\cos\omega t. \label{eq.011000}
\end{eqnarray}
Scolan [1999] applied the Melnikov method to nonlinear ship rolling in waves modeled by
\begin{eqnarray}
 \ddot{x}+2\mu \dot{x}-x(x^2-x_v^2)(x^2-x_l^2)=A\cos(\omega t+\phi). \label{eq.0111000}
\end{eqnarray}
The author provided that the heteroclinic orbits still exist whatever the ``smallness'' of the perturbation as soon as the system is undamped. The
 existence of such  cancellation is otherwise confirmed from an analysis of the erosion of the attraction basin.
 Application of the extended Melnikov's method is used by Wan and Leigh [2008]  for simple single-degree-of-freedom vessel roll motion. In their paper,
 the authors considered the two following ship rolling models: 
\begin{eqnarray}
 &&\ddot{x}+\delta_1 \dot{x}+\epsilon\delta_2\dot{x}|\dot{x}|+x-\alpha x^3=\epsilon f\cos\omega t, \label{eq.01111000}
\end{eqnarray}

\begin{eqnarray}
 &&\ddot{x}+\beta\dot{x}+x(1-x)(1+ax)=F\sin\omega t. \label{eq.011111000}
\end{eqnarray}
The authors compared their results with those obtained by  Falzarano [1990] and Spyrou et al. [2002].
Our aim is to  make a contribution to the study of the transition to chaos in the modified Rayleigh-Duffing oscillator with three well potential 
possessing both homoclinic and heteroclinic orbit by using Melnikov's theory, and  then
 see how the fractal basin boundaries arise and are modified as the damping coefficient is varied. We also focus our attention on the  numerical 
investigation of the strange attractor at parameter values close to the analytically predicted bifurcation curves. 

The paper is organized as follows. In Section 2, we describe and analyze the model.
Section 3 deals with the conditions of existence of Melnikov's chaos resulting from the homoclinic and heteroclinic bifurcation. 
 In Section 4, the basins of attraction of the initial conditions are droped to verify the effectiveness of the method. We provided a  
conclusion and summary in the last section. 

\section{Description  of the model}
 We examine the dynamical transitions in  periodically forced self-oscillating systems
containing the cubic and quintic terms in the restoring force; pure, hybrid quadratic and cubic terms 
in nonlinear damping function as follows:
\begin{eqnarray}
 &&\ddot{x} + \mu (1-\dot{x}^2)\dot{x}+\beta\dot{x}^2+ k_1\dot{x}x+x
+\lambda x^3+\delta x^5=F_0\cos\omega t, \label{eq.1}
\end{eqnarray}
where $\mu, \beta, k_1,  \lambda, \delta, F_0$ and $\omega$ are parameters. Physically, $\mu, \beta$ and $k_1$ represent 
respectively pure cubic and pure, unpure quadratic nonlinear damping
coefficient terms; $\omega$ and $F_0 $ are respectively the  frequency and the amplitude of external periodic forcing. 
 Moreover $ \lambda$ and $\delta$ characterizes the intensity of
 the nonlinearity and $\epsilon$ is the nonlinear damping control parameter. The nonlinear damping term corresponds to the modified Rayleigh
 oscillator, while the nonlinear restoring
force corresponds to the Duffing oscillator, hence the oscillateur is the so-called three-well potential modified Rayleigh-Duffing oscillator.

We derive the fixed points and the phase portrait corresponding to the system (Eq. (\ref{eq.1})) when this is
unperturbed. Eq. (\ref{eq.1}) becomes an unperturbed system and can be rewritten as
\begin{eqnarray}
 \dot{x}=y,\quad  \dot{y}=- x-\lambda x^3 -\delta x^5, \label{eq.2} 
\end{eqnarray}
which corresponds to an integrable Hamiltonian system with the potential function
given by 
\begin{eqnarray}
 V(x) = \frac{1}{2} x^2+\frac{1}{4}\lambda x^4+\frac{1}{6}\delta x^6, \label{eq.3}
\end{eqnarray}
whose associated Hamiltonian function is 
\begin{eqnarray}
 H(x,y)=\frac{1}{2}y^2+ \frac{1}{2}x^2+\frac{1}{4}\lambda x^4+\frac{1}{6}\delta x^6.\label{eq.4}
\end{eqnarray}
From Eqs. (\ref{eq.2})  and (\ref{eq.4}), we can compute the fixed points and analyze their stabilities.

\begin{figure}[ht]
\begin{center}
 \includegraphics[width=4in]{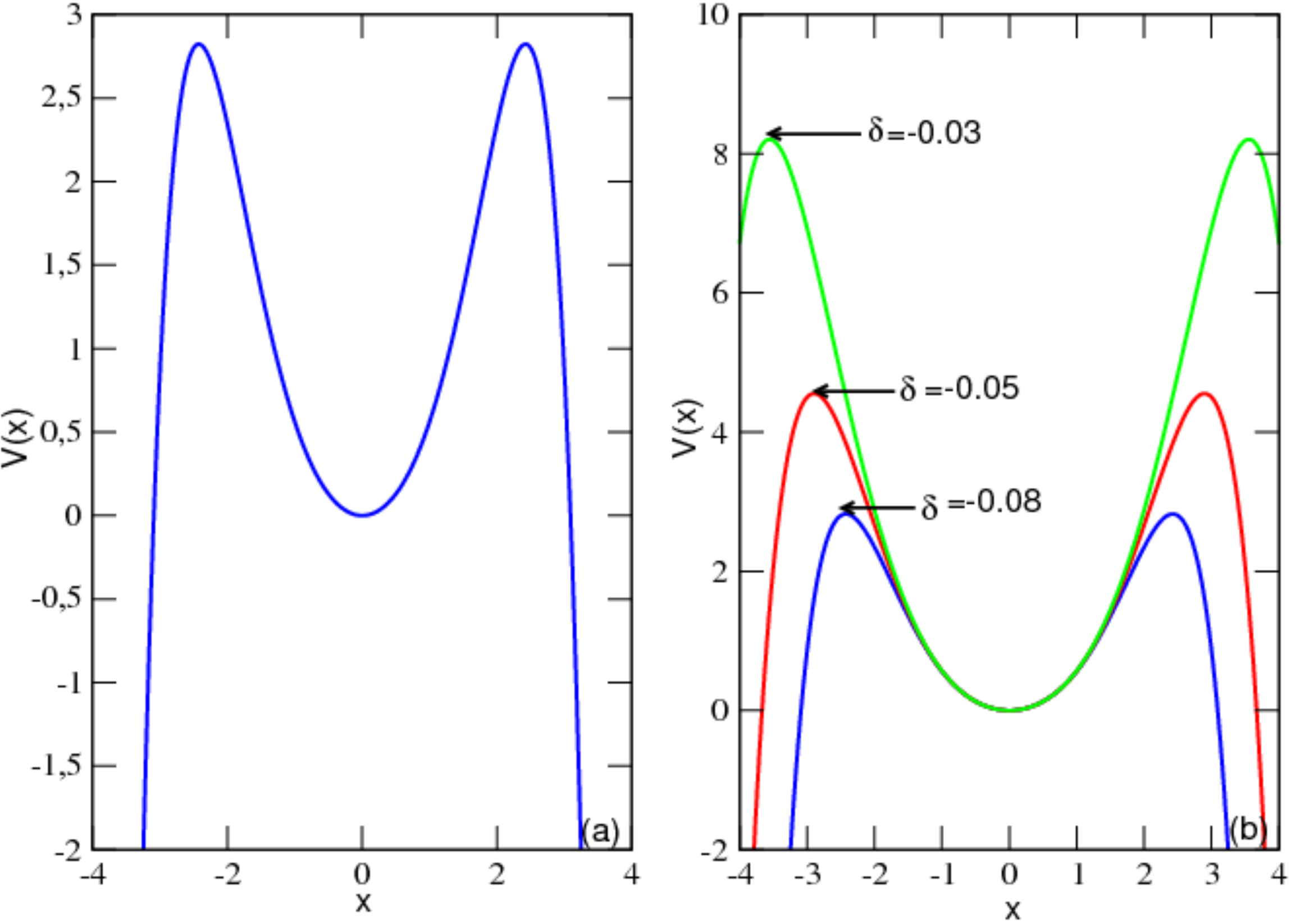}
\end{center}
\caption{$(a):$  Single-well potential of the unperturbed system; 
$(b):$ Shape of single-well potential for different values of $\delta$.}
\label{fig:1}
\end{figure}

$\bullet$ If $\lambda^2<4\delta$, the unperturbed system has only fixed point $(0,0)$ which is a center and the potential $V$ has single well 
(see Fig. \ref{fig:1} ). 

$\bullet$ For $\lambda^2=4\delta$ with   $\lambda>0$ and $\delta<0$, the unperturbed system has three fixed points: two saddles connected
by two heteroclinic orbits and one center. 

\begin{figure}[ht]
\begin{center}
 \includegraphics[width=4in]{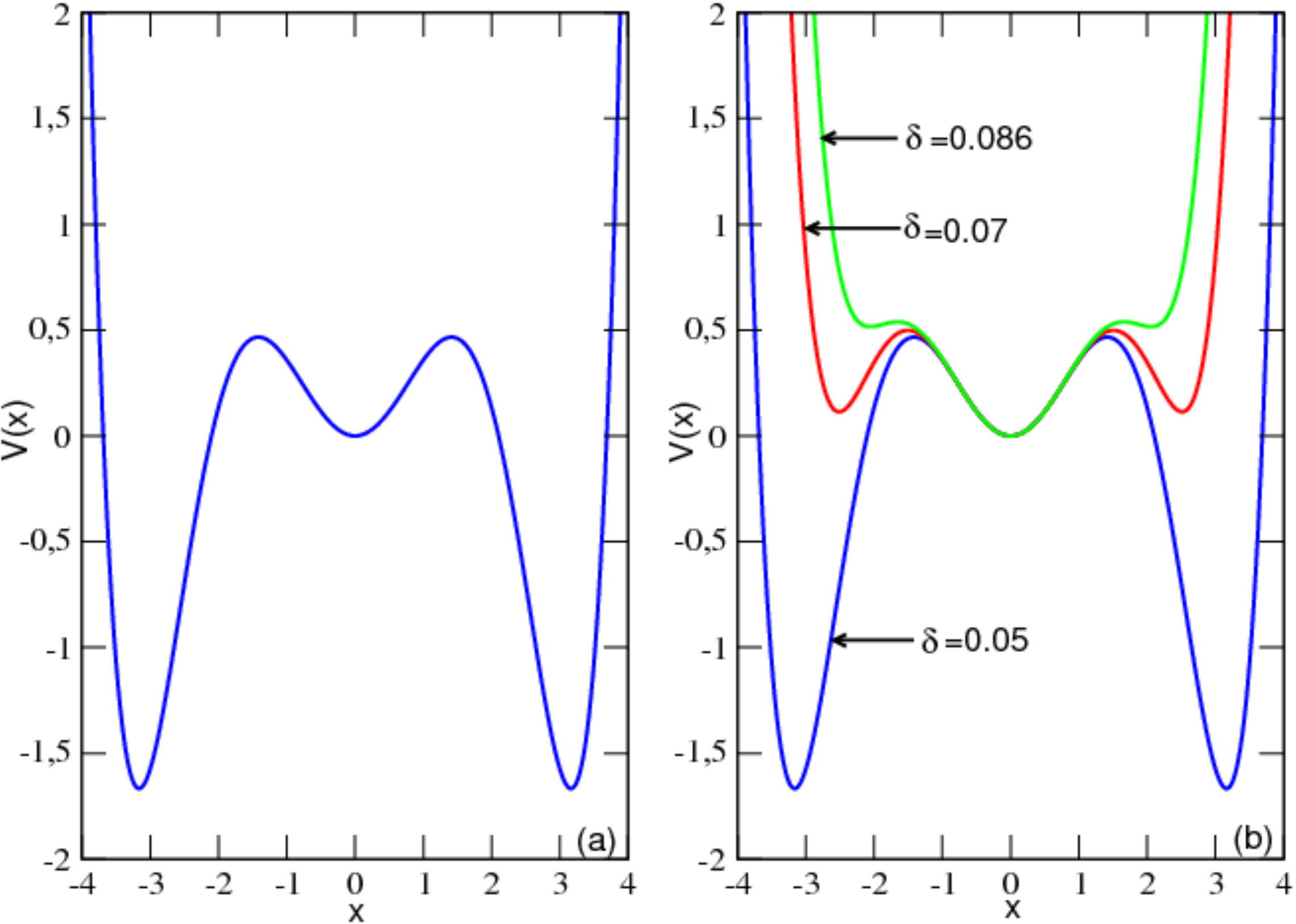}
\end{center} 
\caption{$(a):$ Triple well potential of the unperturbed system; 
$(b):$ Shape of triple well potential for different values of $\delta$.}
\label{fig:2}
\end{figure}

$\bullet$ For $\lambda^2>4\delta$ with   $\lambda<0$ and $\delta>0$, the unperturbed system has  five  fixed points: two saddles connected
by two heteroclinic orbits, and the two saddle points are connected to themselves by one homoclinic orbit. The system has three centers and 
the potential defined by Eq. (\ref{eq.3}) has three-well (see Fig. \ref{fig:2} ).

In this paper, we consider the first and third cases. We set  $\lambda=0.3$, $\delta=-0.08$  and $\lambda=-0.6$, 
$\delta=0.05$ respectively in  single-well potential and three-well potential cases.  

\section{ Melnikov analysis}

In this section, we discuss the chaotic behavior of the system
\begin{eqnarray}
 &&\ddot{x} + \mu (1-\dot{x}^2)\dot{x}+\beta\dot{x}^2+ k_1\dot{x}x+x+\lambda x^3+\delta x^5=
F_0\cos\omega t, \label{eq.5}
\end{eqnarray}
where $\mu, k_1, \beta, k_2, \lambda, \omega$ and $F_0$ are assumed to be small. Hence, we write the dynamical system as
\begin{eqnarray}
 \dot{x}=y,\quad  \dot{y}=- x-\lambda x^3-\delta x^5-\mu (1-\dot{x}^2)\dot{x}-\beta\dot{x}^2+
- k_1\dot{x}x +F_0\cos\omega t. \label{eq.6} 
\end{eqnarray}

Classical Melnikov method is used in many cases to predict the occurrence of chaotic orbits in nonautonomous smooth one-degree-of-freedom 
nonlinear systems [Litak, 2008; Sanju\'an, 1999; Siewe Siewe, 2004; Siewe Siewe, 2005; Soliman, 1992]. It involves transverse intersection of 
stable and unstable manifolds that represent the starting point for
 a successive route to chaotic dynamics. Although a Melnikov theory is merely approximative, it is one of a few methods allowing analytical
 prediction of chaos occurrence. It enables prediction of values of the parameters associated only with the so-called heteroclinic or homoclinic 
chaos. This implies the existence of fractal basin boundaries, and  so-called horseshoes structure of chaos. To deal with such a question, 
we first derive the equation for the separatrix.

When the pertubations are added, the homoclinic or heteroclinic orbits might be broken transversely. And then, by the Smale-Birkoff Theorem 
[ Litak, 2008; Sanju\'an, 1999; Scolan, 1999; Siewe Siewe et al., 2006], horseshose type chaotic
dynamics may appear. It is well known, that the predictions for the appearance of chaos are limited and only valid for orbits starting at points 
sufficiently close to the separatrix. On the other hand it constitutes a first order perturbation method. Although  chaos does not manifest itself
in the form of permanent chaos, and some sorts of transient chaos may showup. However, it manifests itself in terms of fractal basin boundaries as it was 
shown in [Siewe Siewe et al., 2006]. Let us start analysis from the unperturbed Hamiltonian Eq. (\ref{eq.4}). 
At the saddle point $x=0$, for an unperturbed system, the system velocity reaches zero so that the total energy has only 
its potential part. 

We apply the Melnikov method to the system in order to find the necessary criteria for the existence of homoclinic bifurcations and chaos. 
The Melnikov integral is defined as
\begin{eqnarray}
 M(t_0)=\int_{-\infty}^{+\infty}f(x_h,y_h)\wedge g(x_h,y_h)dt, \label{eq.10} 
\end{eqnarray}
where the corresponding differential form $f$ means the gradient of unperturbed hamiltonian while $g$ is a perturbation from Eq. (\ref{eq.6}). 
Eq. (\ref{eq.10}) can be rewritten as follow:
\begin{eqnarray}
 M^\pm(t_0)&=&-\mu\int y_h^2 dt+\mu\int y_h^4dt-k_1\int x_h y_h^2 dt-\beta\int y_h^3 dt+\cr
&&-k_2\int x_hy_h^3 dt+F\int y_h\cos\omega (t+t_0) dt, \label{eq.11} 
\end{eqnarray}
where $t_0$ is the cross-section time of the Poincare map and $t_0$ can be interpreted as the initial time of the forcing term.

Transforming Eqs. (\ref{eq.3}) and (\ref{eq.4}), for a closen nodal energy $(H=0)$  we  get the following 
expression for velocity:
\begin{eqnarray}
 y=\frac{dx}{dt}=\sqrt{2(-\frac{1}{2}x^2-\frac{\lambda}{4}x^4-\frac{\delta}{6}x^6)}. \label{eq.7}
\end{eqnarray}
Now one can perform the integration over $x$:
\begin{eqnarray}
 t-t_0=\pm\int\frac{dx}{x\sqrt{-1-\frac{\lambda}{2}x^2-\frac{\delta}{3}x^4}}.\label{eq.8} 
\end{eqnarray}

\subsection{A single well potential case}

In this case, the fixed points are connected by an  the heteroclinic trajectories given by [Siewe Siewe et al., 2004]
\begin{eqnarray}
&& x_h=\pm \frac{\alpha\gamma\sinh\Omega t}{\sqrt{2[1+(1-\gamma^2)\cosh{\Omega(t-t_0)]}}},\cr
 &&y_h=\pm\frac{\alpha\gamma\Omega\cosh\Omega t}{\sqrt{2[1+(1-\gamma^2)\cosh{\Omega(t-t_0)]^3}}}, \label{eq.9} 
\end{eqnarray}

where 
\begin{eqnarray}
&& \Omega=x_2^2\rho\sqrt{\frac{-\delta(1+\rho^2)}{2}};\quad \alpha^2=x_2^2(\rho^2+3);\quad \gamma^2=\frac{2\rho^2}{3(1+\rho^2)};\cr 
&&\rho^2=\frac{\lambda+\sqrt{\lambda^2-4\delta}}{-\lambda+\sqrt{\lambda^2-4\delta}}. \nonumber
\end{eqnarray}

 Note that the central
saddle point $x_0=0$ is reached in time $t$ corresponding to $+\infty$ and $-\infty$ respectively.

After some algebra, the Melnikov function can thus be evaluated and a necessary condition for the onset of Melnikov chaos is given by

\begin{eqnarray}
F\ge \frac{\alpha^3\Omega^3\sqrt{2(1-\gamma^2)}\sinh(\frac{\pi\omega}{2\Omega}) }{64\pi\omega\gamma}\left[\mu(\Upsilon_0+\Upsilon_1)+ 
\frac{ 8\alpha^3\gamma^3\Omega^2\beta(7-6\gamma^2)}{105(1-\gamma^2)^{3/2}}
\right],
\end{eqnarray}

where 
\begin{eqnarray}
\Upsilon_0=\frac{\alpha^2\Omega}{4}\left(\frac{3\gamma^2-1}{1-\gamma^2}+\frac{3\gamma^2+1}{2\gamma}\ln\frac{1+\gamma}{1-\gamma}\right) \nonumber
\end{eqnarray}
and
\begin{eqnarray}
&\Upsilon_1=\frac{\alpha^4\Omega^3}{1280\gamma^2}\times \cr
&\left(\frac{315\gamma^8-420\gamma^6+38\gamma^4+20\gamma^2+15}{(1-\gamma^2)^2}+
\frac{15(3\gamma^2+1)(1+7\gamma^4)}{2\gamma}\ln\frac{1+\gamma}{1-\gamma}\right).\nonumber
\end{eqnarray}

\begin{figure}[ht]
\begin{center}
 \includegraphics[width=4in]{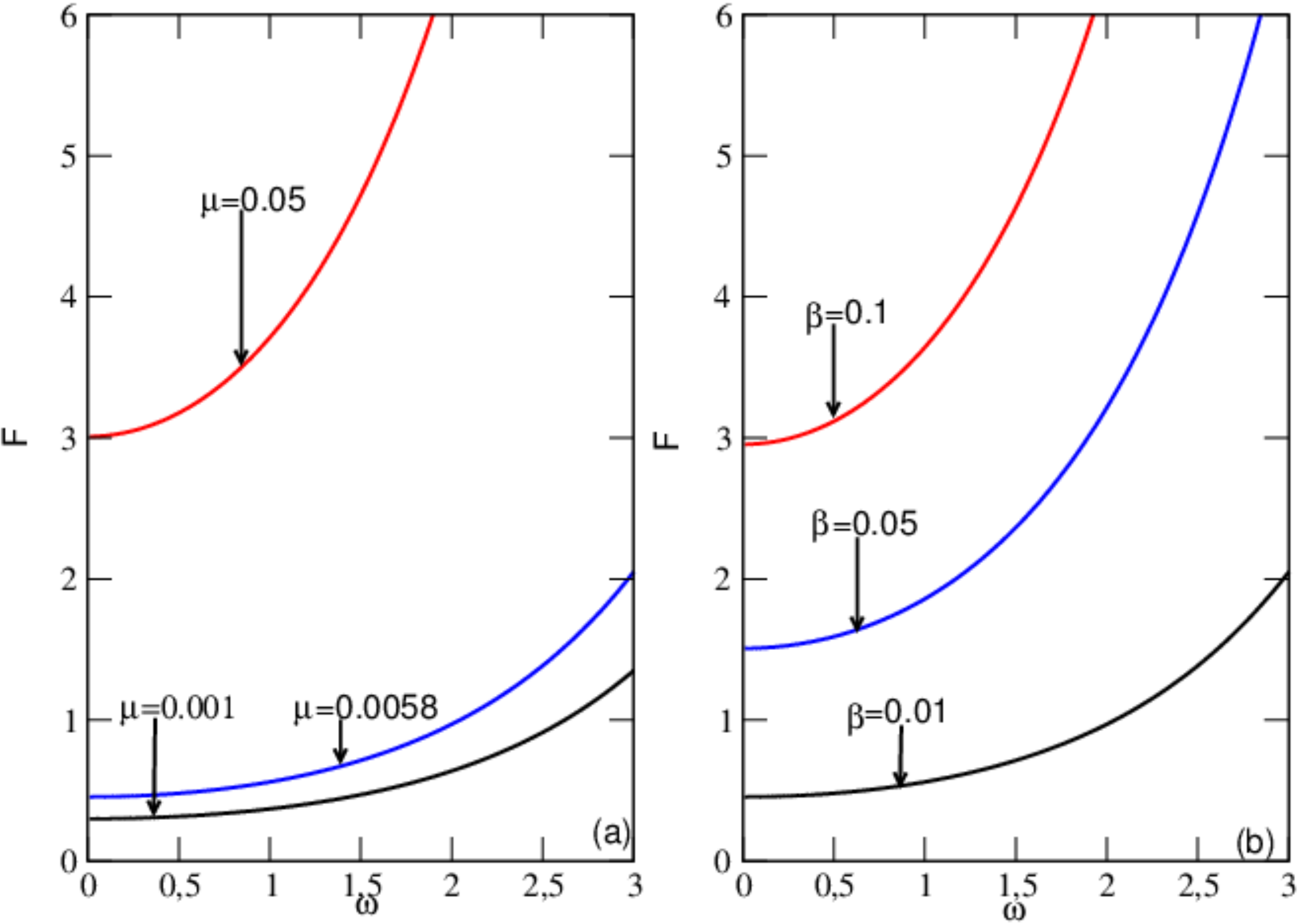}
\end{center}
\caption{ The critical curves for chaotic motions of a modified Rayleigh-Duffing oscillator in a single well potential case in  $(\omega, F)$ plane with
$\lambda=0.3$, $\delta=-0.08$; $(a):$ effect of pure cubic damping coefficient $\mu$ and  $(b):$ effect of pure quadratic damping coefficient $\beta$.}
\label{fig:3a}
\end{figure}

Figs.\ref{fig:3a} $(a), (b)$ show the critical external forcing amplitude for different values of $\mu$ and $\beta$ respectively.
 The region under the curve represents the domain leading to suppression of horseshoes chaos. It appears that, depending on 
the parameters of the systems, the structure could or could not present chaotic dynamics. 

\subsection{A three wells potential case}

In this subsection, the three wells type (the bounded double hump potential: Fig.\ref{fig:2}), the system
Eq. (\ref{eq.2}) also has two hyperbolic fixed points and each point process two different types of orbit: a heteroclinic
orbit connecting the two saddle points an defined as [Siewe Siewe et al., 2005]:

\begin{eqnarray}
&& (x_{he},y_{he}) =\left(\pm \frac{x_1\sqrt{2}\cosh(\frac{\theta t}{2})}{\sqrt{(\Psi+\cosh{\theta t)}}},\quad \mp\frac{\theta x_1\sqrt{2}(1-\Psi)
\sinh(\frac{\theta t}{2})}{2\sqrt{(\Psi+\cosh{\theta t)^3}}}\right) \label{eq.t1} 
\end{eqnarray}

and a symmetric pair of homoclinic trajectories connected each point to itself given by

\begin{eqnarray}
&& (x_{ho},y_{ho}) =\left(\pm \frac{x_1\sqrt{2}\sinh(\frac{\theta t}{2})}{\sqrt{(-\Psi+\cosh{\theta t)}}},\quad \pm\frac{\theta x_1\sqrt{2}(1-\Psi)
\cosh(\frac{\theta t}{2})}{2\sqrt{(-\Psi+\cosh{\theta t)^3}}}\right), \label{eq.t2} 
\end{eqnarray}
where 
\begin{eqnarray}
 &\theta=x_1^2\sqrt{2\delta(\rho^2-1)}{2};\quad \alpha^2=x_2^2(\rho^2+3);\quad \Psi=\frac{5-3\rho^2}{3\rho^2-1};\cr
&\rho^2=\frac{\lambda+\sqrt{\lambda^2-4\delta}}{-\lambda+\sqrt{\lambda^2-4\delta}}, \nonumber
\end{eqnarray}
and
\begin{eqnarray}
 x_1=\sqrt{\frac{-1}{2\delta}(\lambda+\sqrt{\lambda^2-4\delta})}, \nonumber
\end{eqnarray}

\begin{eqnarray}
 x_2=\sqrt{\frac{-1}{2\delta}(\lambda-\sqrt{\lambda^2-4\delta})}. \nonumber
\end{eqnarray}

Let us first consider the case of heteroclinic orbit. After substituting the equations of the heteroclinic orbits $x_h$ and $y_h$ given 
in Eq. (\ref{eq.t1}) into 
Eq. (\ref{eq.11}), we calculate the
Melnikov function and a sufficient condition for the appearance of chaos in sense of Smale is given by

\begin{eqnarray}
 F\ge \frac{\theta\sinh(\frac{\pi\omega}{\theta})}{2\pi\omega x_1}\left[\frac{2(1+\Psi)\mu }{x_1^2\theta\chi_{he}}+
 \frac{ 2x_1^3\theta^2(13-\Psi)\beta}{105(1-\Psi)} \right],
\end{eqnarray}

where 
\begin{eqnarray}
\chi_{he}=2+\Psi+\frac{\theta^2x_1^2\Xi_{he}}{160(1+\Psi)^3}+
\frac{\arcsin\Psi+\frac{\pi}{2}}{\sqrt{1-\Psi^2}} \left[1+2\Psi+
\frac{\theta^2x_1^2\Delta_{he}}{160(1+\Psi)^3}\right], \nonumber
\end{eqnarray}
with
\begin{eqnarray}
\Xi_{he}=\frac{402\Psi^4+4231\Psi^3+5063\Psi^2+4448\Psi+976}{3(\Psi-1)} \nonumber
\end{eqnarray}
and
\begin{eqnarray}
\Delta_{he}=\frac{2082\Psi^4+3696\Psi^3+5889\Psi^2+405\Psi+135}{3(\Psi-1)}. \nonumber
\end{eqnarray}

\begin{figure}[ht]
\begin{center}
 \includegraphics[width=4in]{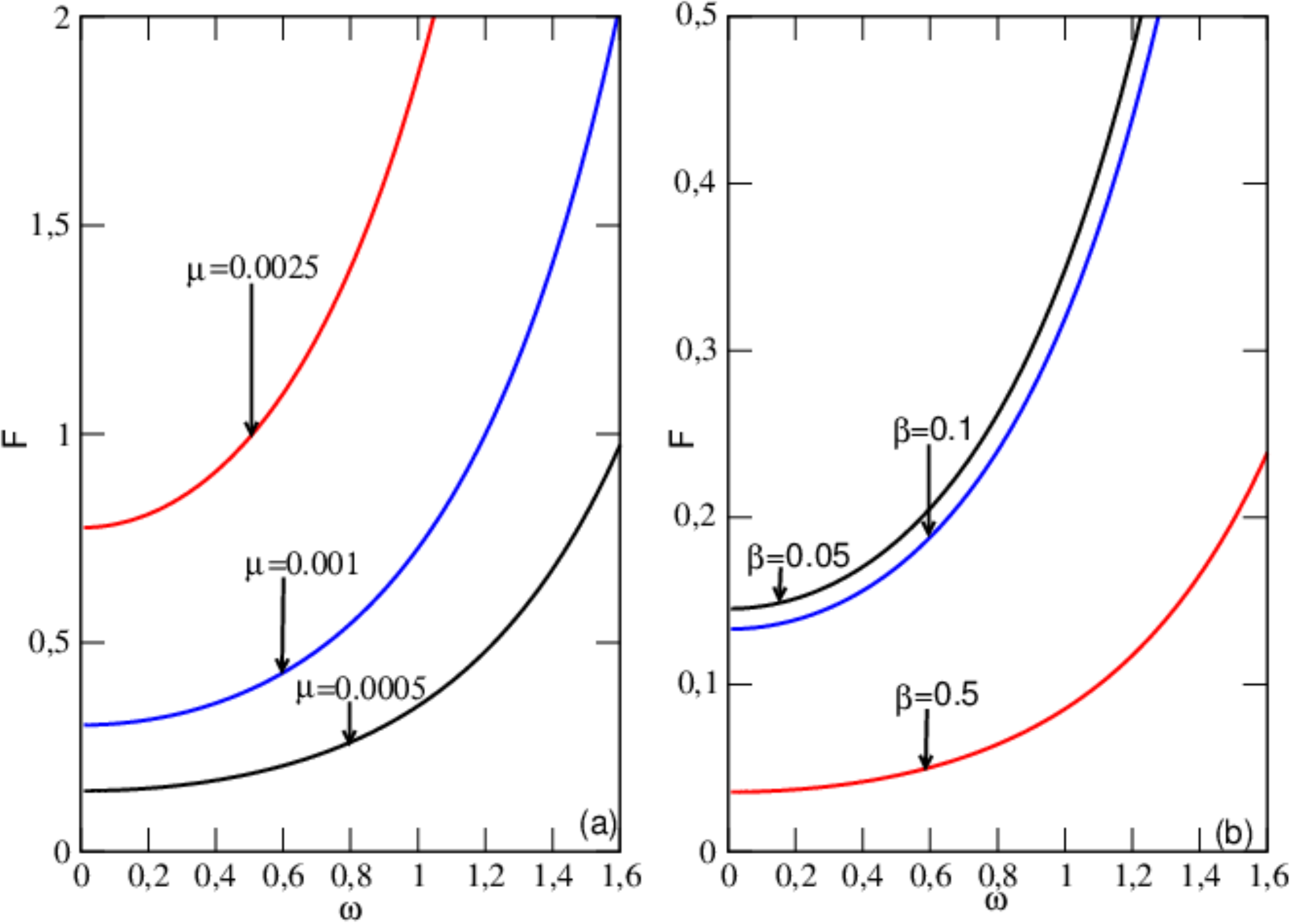}
\end{center}
\caption{ The critical curves for heteroclinic chaos of a modified Rayleigh-Duffing oscillator in a single well potential case in  $(\omega, F)$ plane with
$\lambda=-0.6$, $\delta=0.05$; $(a):$ effect of pure cubic damping coefficient $\mu$ and  $(b):$ effect of pure quadratic damping coefficient $\beta$.}
\label{fig:3}
\end{figure}

From each of these relations, the threshold values of the parameters for which the stable and unstable manifolds to the heteroclinic point intersect 
are obtained and the complicated behavior occurs whenever the above condition is realized.
We study the chaotic threshold as a function of only the frequency parameter $\omega$. A typical plot is shown in  Fig. \ref{fig:3}, 
in which the critical heteroclinic bifurcation curves are plotted  versus the frequency parameter $\omega$.
Figs.\ref{fig:3} $(a), (b)$ show the critical external forcing amplitude for different values of $\mu$ and $\beta$ respectively. 
One can see (Fig.\ref{fig:3} $(a)$) that when the value of the cubic damping parameter $\mu$ increases, the thresholds of the critical values 
for heteroclinic bifurcation of the harmonic excitation $F$ increase. The contrary effect is observed with the pure quadratic damping coefficient
 $\beta$ (Figs.\ref{fig:3} $(b)$). We concluded  that the parameters $F$ and $\mu$
have the same effect on the critical value for chaotic motions which is the opposit one of parameter $\beta$.

Now, consider the case of homoclinic orbit. Substituting  Eq. (\ref{eq.t2}) into Eq. (\ref{eq.11}),  we calculate
the Melnikov function. It is known that the intersections of the homoclinic orbits are the necessary conditions for the
existence of chaos. The Melnikov function theory
measures the distance between the perturbed stable and unstable manifolds in the Poincar\'e section. If $M^\pm(t0 )$ has a simple zero,
 then a homoclinic bifurcation occurs, signifying the possibility of chaotic behavior. This means that only necessary conditions for the appearance
 of strange attractors are obtained from the Poincar\'e-Melnikov-Arnold analysis, and therefore one always has the chance of
finding the sufficient conditions for the elimination
of even transient chaos. Then the necessary condition for which the invariant manifolds intersect is given by

\begin{eqnarray}
 F\ge\frac{\theta}{2 x_1 \sin(\frac{2\omega}{\theta}) }\left[ \frac{ x_1^2\mu\theta\chi_{ho}}{4(1+\Psi)^2} 
+ \frac{ 2\sqrt{2}k_1}{15(1+\Psi)^2}\right],
\end{eqnarray}
where 
\begin{eqnarray}
\chi_{ho}=\Psi^2+3\Psi-2-\frac{\theta^2x_1^2\Xi_{ho}}{120(1+\Psi)^2}+ \frac{\arcsin\Psi-\frac{\pi}{2}}{\sqrt{1-\Psi^2}}\times\cr
\left[\Psi(3\Psi-1)- \frac{\theta^2x_1^2\Delta_{ho}}{120(1+\Psi)^2}\right]; \nonumber
\end{eqnarray}
with
\begin{eqnarray}
\Xi_{ho}=\frac{5\Psi^5+42\Psi^4+603\Psi^3+561\Psi^2+566\Psi+112}{\Psi-1}; \nonumber
\end{eqnarray}
and
\begin{eqnarray}
\Delta_{ho}=\frac{320\Psi^4+400\Psi^3+765\Psi^2+315\Psi+90}{\Psi-1}. \nonumber
\end{eqnarray}

\begin{figure}[ht]
\begin{center}
\includegraphics[width=4in]{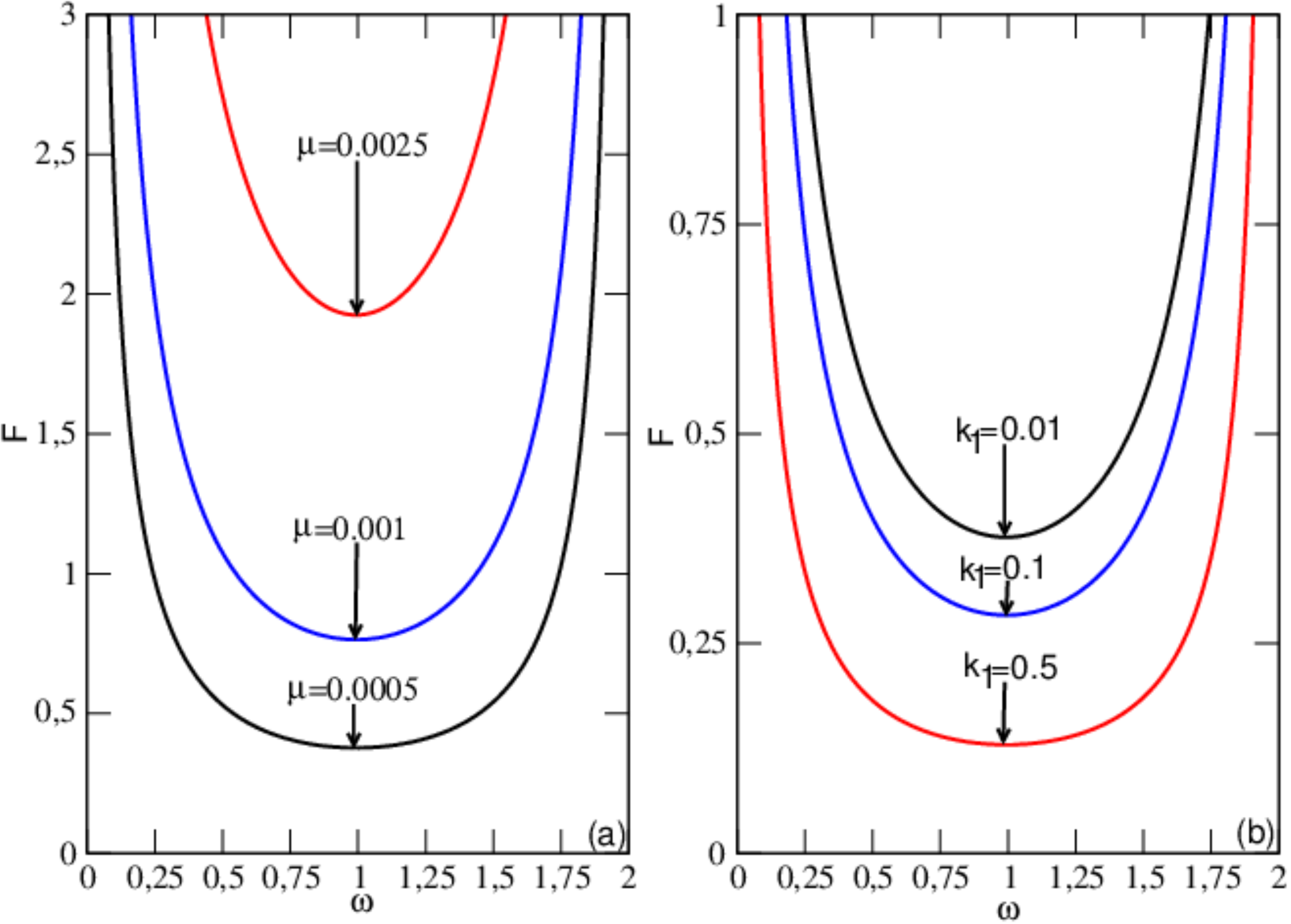}
\end{center}
\caption{ The critical curves for homoclinic chaos of a modified Rayleigh-Duffing oscillator in a triple well potential case in  $(\omega, F)$ plane with
$\lambda=-0.6$, $\delta=0.05$; $(a):$ effect of pure cubic damping coefficient $\mu$ and  $(b):$ effect of unpure quadratic damping coefficient $k_1$ }
\label{fig:4}
\end{figure}

These conditions provided a domain of the parameters  where the system has transverse homoclinic orbits resulting in possible chaotic behavior. In
order to have a visual information, we have plotted in Figs.\ref{fig:4} $(a), (b)$ the dependence of the external excitation $F$ on the frequency $\omega$ 
for different values of $\mu$ and $k_1$ respectively. These figures, show that
the parameters   $\mu$ and  $k_1$ have the  opposit  effect on the homoclinic critical value for chaotic motions. 

 \section{Numerical simulations and analysis}

\begin{figure}[ht]
\begin{center}
 \includegraphics[width=5in]{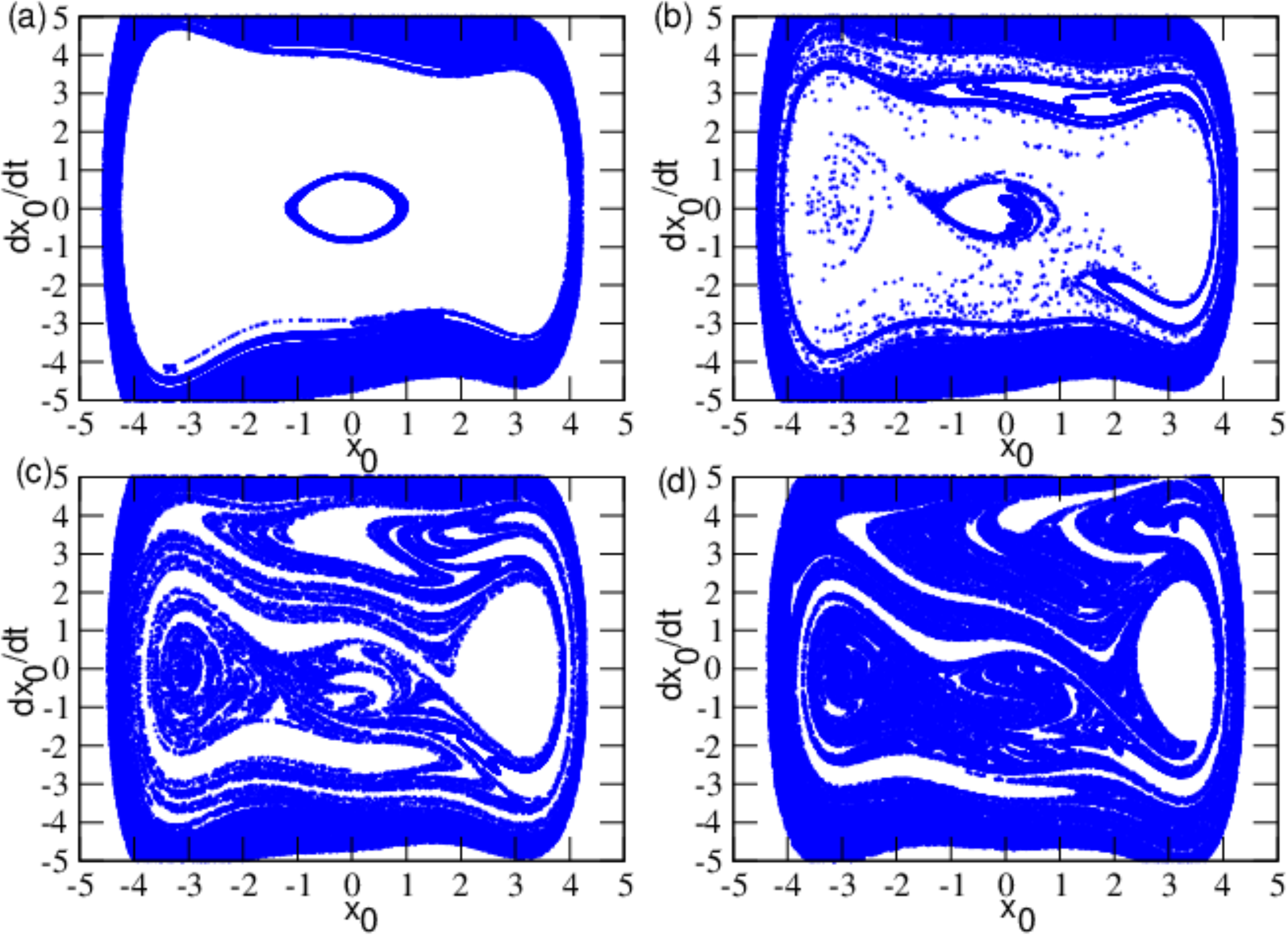}
\end{center}
\caption{ Basin of attraction in heteroclinic  chaos  of a modified Rayleigh-Duffing oscillator in a triple well potential: $(a) F= 0.05, (b) F= 0.1, 
(c) F= 0.4$ and $(d) F= 0.85$, the others parameters are  $\lambda=-0.6, \delta=0.05, k_1=0.05, \beta=0.05, \mu=0.0005, \omega=1$.}
\label{fig:b41}
\end{figure}

\begin{figure}[ht]
\begin{center}
 \includegraphics[width=5in]{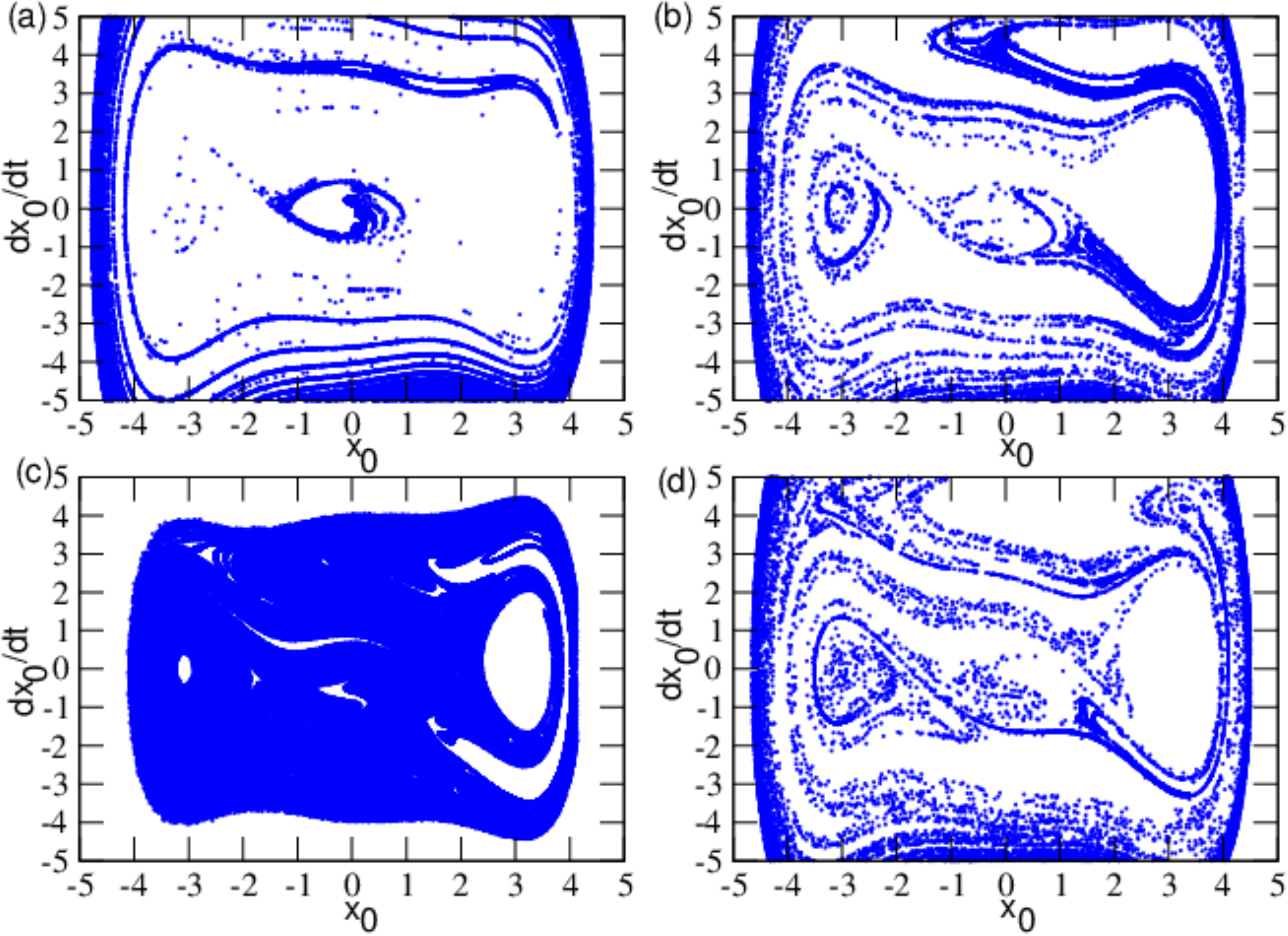}
\end{center}
\caption{ Basin of attraction in homoclinic chaos  of a modified Rayleigh-Duffing oscillator in a triple well potential: $(a) F= 0.1, (b) F= 0.4, 
(c) F= 0.5$ and $(d) F= 0.85$, the others parameters are  $\lambda=-0.6, \delta=0.05, k_1=0.1, \beta=0.05, \mu=0.0005, \omega=1$.}
\label{fig:b42}
\end{figure}

\begin{figure}[ht]
\begin{center}
\includegraphics[width=4in]{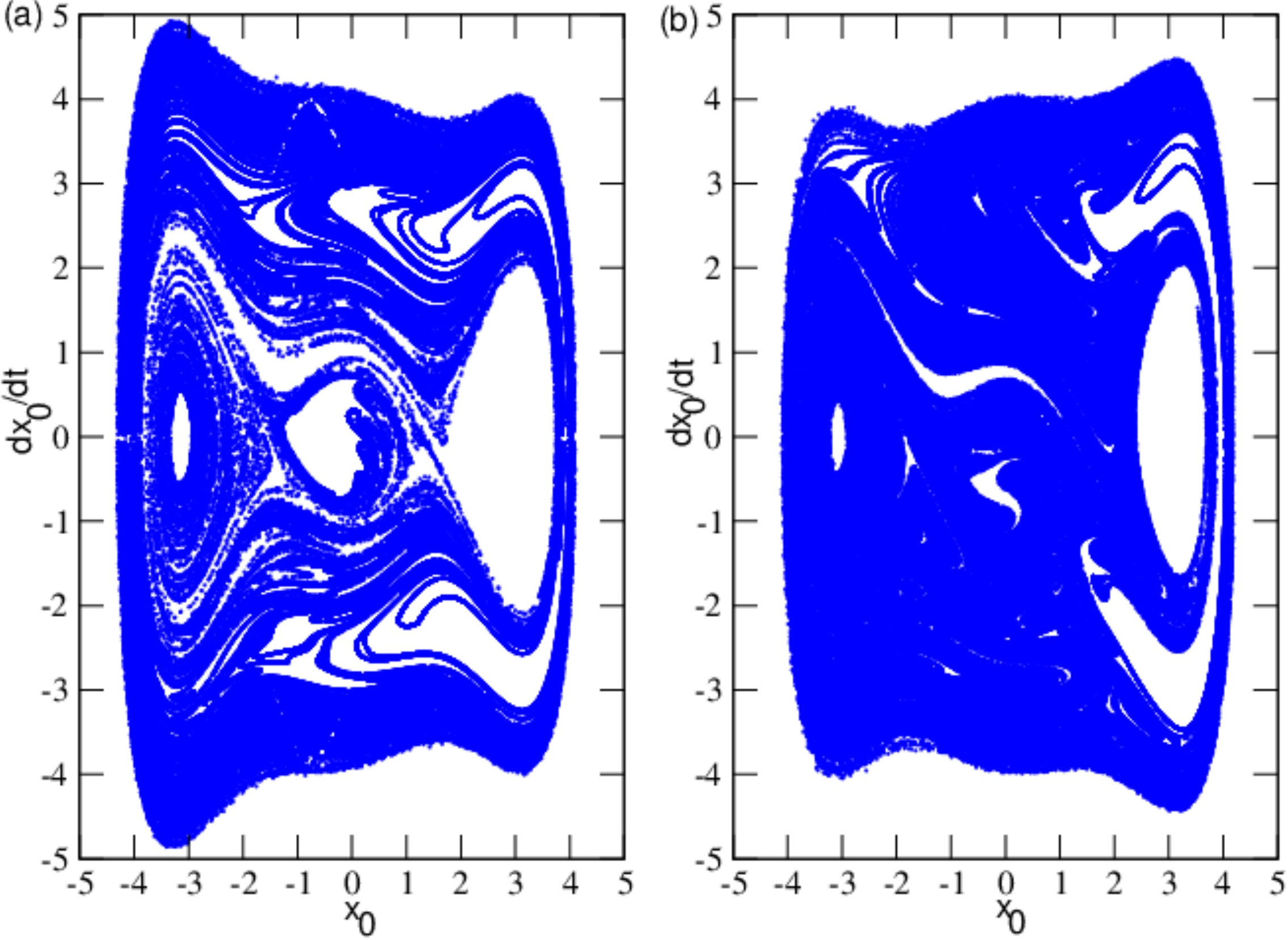}
\includegraphics[width=4in]{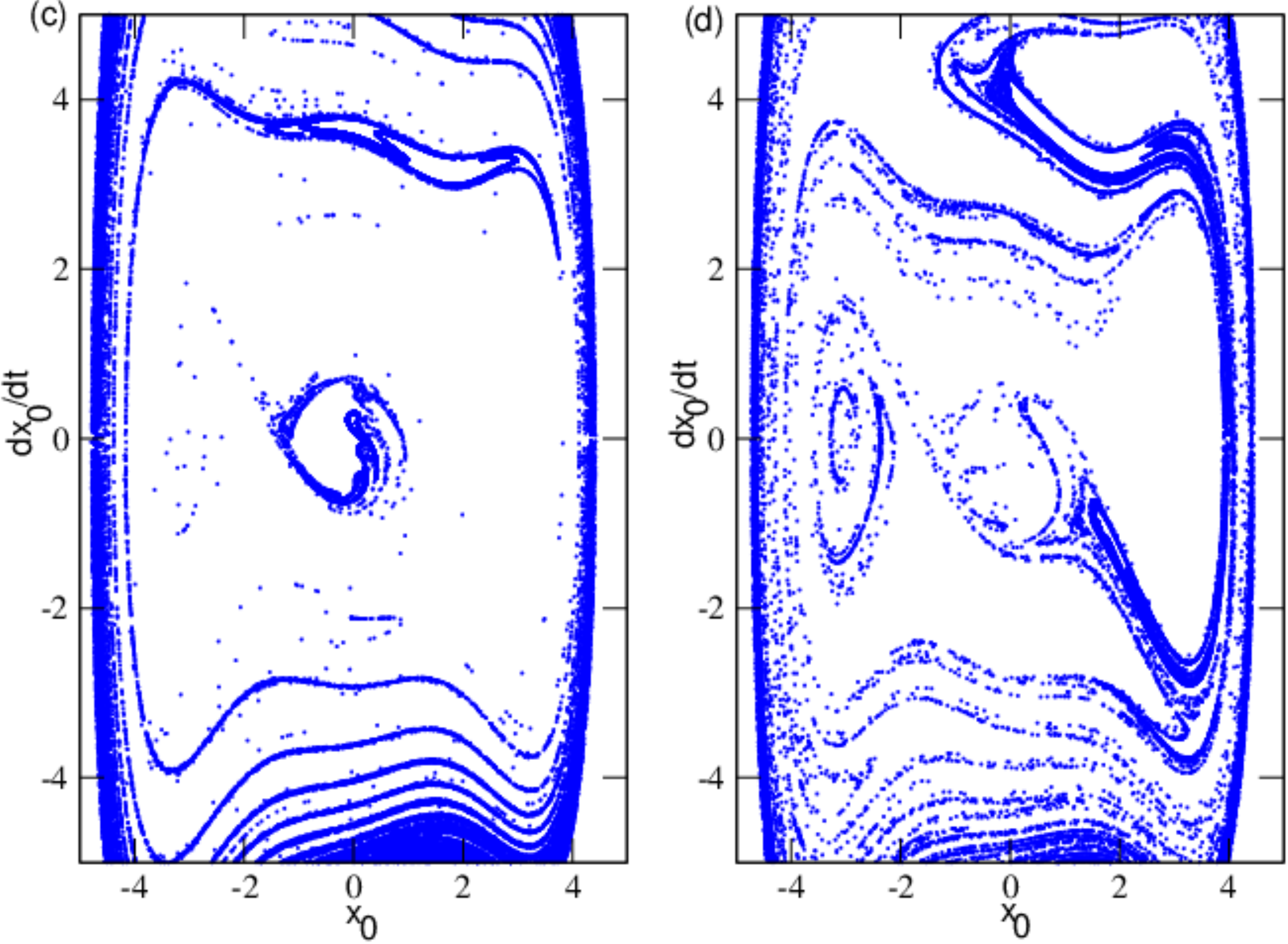}
\includegraphics[width=4in]{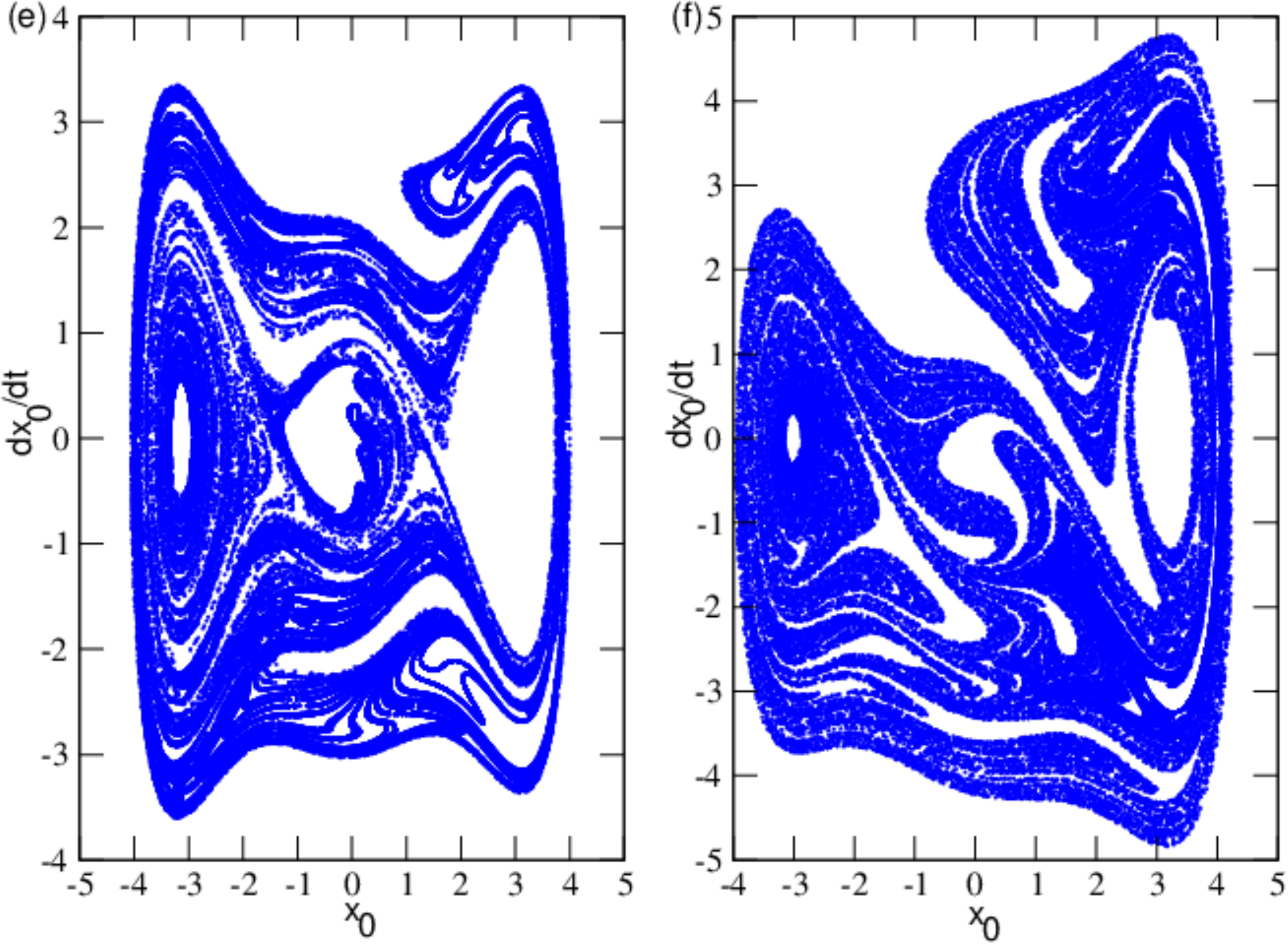}
\end{center}
\caption{ Effects of nonlinear damping parameters on basin of attraction in the triple well potential $(a)$ and  $(b)$ effect of $F$, 
$(c)$ and  $(d)$ effects of $k_1$  and
$(e)$ and  $(f)$ effects of $\mu$.}
\label{fig:b43}
\end{figure}

  In order to verify the analytical results obtained in the previous sections, we have numerically integrated the system
 by using a fourth order Runge-Kutta algorithm to investigate the heteroclinic and the homoclinic chaos in the model. We consider the single well and 
triple well cases. We checked 
 the effect of the nonlinear damping terms on the motion of the oscillator. In the same vein, we investigated how the basins of attraction are
 affected as the parameter $\mu, \beta, k_1$ and $F$ varied. We  see through 
 Figs.\ref{fig:b41}, \ref{fig:b42} and \ref{fig:b43} that the basin boundaries become fractal or basin of attraction is destroyed. 
The blue zone stands for the area where the choice of the initial conditions leads to a chaotic
 motion while the white area is the  domain of periodic or quasi-periodic oscillations.
  This means that
  the damping parameter  $\mu, \beta, k_1$ and amplitude of external forced $F$ have contributed to
These numerical results obtained confirm exactly 
 the analytical results.

%
%
%
%

\section{Discussion and conclusions}

In this paper, the dynamics of a forced nonlinear damped system, the $\phi^6$ modified Rayleigh-Duffing oscillator  has
been studied. The analytical criteria for the appearance of chaos in the sense of Smale have also been derived using Melnikov theory.
The effects of nonlinear damping and amplitude of external forced on Melnikov critical values have been investigated. 
In the single well potential case, we noticed that the nonlinear damping parameters $\mu$ and $\beta$ and the forced amplitude $F$ have the same effects.
Melnikov critical value increases with each of these parameters. The parameters $\mu$ and $F$ have the same effects on homoclinic  and 
heteroclinic Melnikov criterion in three well potential case but the quadratic damping $\beta$ and $k_1$ have the contrary effects. It should be
 noted that the pure quadratic damping affected the heteroclinic bifurcation and unpure quadratic damping
affected the homoclinic bifurcation. A convenient demonstration of the use and accuracy of
the method, is obtained from the basin of attraction.
 By means of the basin of attraction, we have shown that for certain regions of parameter space, the deterministic system driven harmonically
experience behaviors that may be chaotic or non-chaotic. 

Finally, the above assessment reveals that roll damping is a very critical parameter in motion characteristics of a ship. Therefore estimation of roll
damping may lead to accurate prediction of chaotic ship rolling  motions. In principle, it may be explained as the influence between nonlinear damping 
and excitation force.

\section*{Acknowlegments}
The authors thank IMSP for financial support. We also thank 
Professor Paul Woafo, Professor Rom\'eo Nana Nbendjo an Doctor  herv\'e Enjieu Kadji for their suggestions and collaboration.


\end{document}